\documentstyle[prl,aps,twocolumn,epsfig]{revtex}

\begin{document}
%______________________________________________________________________ TITLE

\title{Optimal State Discrimination Using Particle Statistics}

\author{S. Bose$^{1,2}$, A. Ekert$^{3}$, Y. Omar$^{4,5}$, N. Paunkovi\'c$^{4}$ and V. Vedral$^{6}$}

\address{$^{1}$ Institute for Quantum Information, California
Institute of Technology, CA 91125\\ $^{2}$ Department of Physics
and Astronomy, University College London, Gower Street,
London WC1E 6BT, United Kingdom\\
$^{3}$ Centre for Quantum Computation, DAMTP, University of
Cambridge, Cambridge CB3 0WA, United Kingdom\\
$^{4}$ Centre for Quantum Computation, Clarendon Laboratory,
University of Oxford,\\ Parks Road, Oxford OX1 3PU, United
Kingdom\\
$^{5}$ Centro de F\'isica de Plasmas, Instituto Superior
T\'ecnico, P-1049-001 Lisbon, Portugal\\
 $^{6}$Optics Section, The Blackett Laboratory, Imperial
College, London SW7 2BZ, United Kingdom}

\date{September 8, 2003}

\maketitle

%______________________________________________________________________ ABSTRACT

\begin{abstract}
We present an application of particle statistics to the problem of
optimal ambiguous discrimination of quantum states. The states to
be discriminated are encoded in the internal degrees of freedom of
identical particles, and we use the bunching and antibunching of
the external degrees of freedom to discriminate between various
internal states. We show that we can achieve the optimal
single-shot discrimination probability using only the effects of
particle statistics. We discuss interesting applications of our
method to detecting entanglement and purifying mixed states. Our
scheme can easily be implemented with the current technology.
\end{abstract}

\pacs{Pacs No: 03.67.-a, 03.65.-w, 05.30.-d}

%______________________________________________________________________

Recently, there has been an emerging interest in the use of
particle statistics (both bosonic and fermionic) for quantum
information processing \cite{omar,paunkovic}. In fact, it was
shown that a useful task such as entanglement concentration could
be accomplished, even if non-optimally, using only the effects of
quantum statistics, without the need for any other interactions
\cite{paunkovic}. The above investigations differ significantly
from some previous suggestions, where either anyonic statistics
\cite{kitaev} or the effects of electronic statistics in
conjunction with other interactions \cite{divincenzo-loss,costa}
were used for quantum information processing. Schemes using only
particle statistics \cite{omar,paunkovic} would be very useful for
tasks implemented with identical particles that interact very
weakly or not at all with each other, such as photons or neutrons.
This weak interaction can be beneficial for information processing
as it may reduce the unwanted coupling to the environment. Such
schemes are also extremely general in the sense of being
independent of the actual particle species. It is however not
known whether such schemes can accomplish quantum information
processing {\em efficiently}. Here we present a particular quantum
information processing task involving two qubits and show that it
can be performed {\em optimally} using {\em only} quantum
statistics. Moreover, we point out how the task of discriminating
quantum states can be applied to detecting entanglement and
purifying mixed states. We also discuss how to generalize these
tasks to $N$ qubits and argue that quantum statistics could be
used to perform even this generalized task optimally. While the
two qubits and other small $N$ qubits versions of our protocol can
be tested with photons, electrons, neutrons or atoms, the large
number of qubits versions could have interesting implementations
in optical lattices \cite{opt-latt,jaksch}.

One of the striking aspects of quantum mechanics is that it is not
possible to perfectly discriminate between two states unless they
are orthogonal. Suppose someone prepares two qubits encoded in the
internal degrees of freedom of two identical particles --- say, in
the spin of two electrons or the polarization of two photons ---
in one of the following two possible states:
\begin{itemize}
\item spins aligned (parallel) and pointing in an arbitrary
direction;

\item spins anti-aligned and pointing in an arbitrary direction.

\end{itemize}
We will assume that the two states are equally likely for the sake
of simplicity, but all the presented results will be valid for any
\emph{a priori} distribution. Note also that, unless otherwise
stated, whenever we mention spin we will actually refer to any
two-dimensional internal degree of freedom, be it for fermions or
for bosons.  As we do not have any knowledge of the direction of
alignment in both of the above cases, the overall states are
mixed. They are described respectively by the following density
operators:
\begin{equation}
\rho_2=\frac{1}{4\pi}\int d\Omega
|\Omega\rangle\langle\Omega|\otimes|\Omega\rangle\langle\Omega|,
\label{rhostate}
\end{equation}
and
\begin{equation}
\sigma_2=\frac{1}{4\pi}\int d\Omega
|\Omega\rangle\langle\Omega|\otimes|\Omega^\bot\rangle\langle\Omega^\bot|.
\end{equation}
Here, the subscripts indicate that we are considering two
particles. The ket $|\Omega\rangle$ represent the spin-up state
along the axis defined by the angle $\Omega$, while
$|\Omega^\bot\rangle$ is the orthogonal state spin-down. Thus,
$\rho_2$ represents an equal mixture of spins aligned along an
arbitrary axis in space, while $\sigma_2$ is the equal mixture of
anti-aligned spins of two spin-$\frac{1}{2}$ particles. It is
impossible to discriminate between these states perfectly because
they are not orthogonal.

Optimal results are known for the discrimination of any two given
quantum states $\eta$ and $\eta^{\prime}$ \cite{hels}. The maximal
probability of ambiguously discriminating between two \emph{a
priori} equally likely quantum states in a single-shot measurement
is given by the Helstrom formula:
\begin{equation}
P_H(\eta,\eta^{\prime})=\frac{1}{2}+\frac{1}{4}Tr|\eta
-\eta^{\prime}|.
\end{equation}

We now present a procedure for discriminating between $\rho_2$ and
$\sigma_2$, both for fermions and bosons, based only on the
effects of particle statistics. To use these effects we interfere
particles at a beam splitter. Here we use \emph{beam splitter} in
a generic sense, referring not only to the common optical element
(partially silvered mirror) used with photons, but also to any
device presenting an analog behavior for other kinds of particles,
as already suggested for electrons \cite{yamamoto}. In order to
distinguish the states we rely on path measurements that
discriminate between bunching and antibunching, and that are
performed on our particles after letting them pass simultaneously
through a $50/50$ beam splitter. Note that in such balanced beam
splitters two indistinguishable particles will always bunch if
they are bosons \cite{mandel}, and always antibunch if they are
fermions \cite{yamamoto} (see also \cite{loudon}).

For fermions, our guess in the case of the antibunching result is
that the input state was $\rho_2$, while in the case of bunching
is that it was $\sigma_2$. The probability of success of our
procedure is then:
\begin{equation}
P_{BS} (\rho_2, \sigma_2)=\frac{3}{4}. \label{probbs}
\end{equation}
This probability can easily be calculated by noticing that the
only case for which our guess could be incorrect is when we have
the antibunching result. Then, we conclude that the input state
was $\rho_2$, while it could have actually been $\sigma_2$. On the
other hand, in the case of the bunching result we know for sure
that the input state was $\sigma_2$, since --- according to the
Pauli exclusion principle --- two particles with aligned spins
cannot end up in the same output arm of the beam splitter. When
our input state is $\sigma_2$, the antibunching happens with
probability $1/2$, giving in total a probability of incorrect
inference of $1/4$. In the case of bosons, our protocol is exactly
the opposite of the fermionic one, but yields precisely the same
efficiency: this time the antibunching results stand for
$\sigma_2$, whereas the bunching ones stand for $\rho_2$, but the
probability of success coincides with Eq. (\ref{probbs}).

Interestingly, the Helstrom formula gives the same result for the
maximal probability of discriminating between these two states:
\begin{equation}
P_{H} (\rho_2, \sigma_2) =\frac{3}{4}.
\end{equation}
We can thus conclude that our procedure is optimal for both
fermions and bosons.

Helstrom's probability can still be achieved in other cases using
the effects of quantum statistics. We now introduce a case of
special interest that can be applied to other tasks such as
entanglement detection and state purification, as will be shown
later in the paper. Suppose that we have to discriminate between
the following two states:
\begin{itemize}
\item spins aligned (parallel) and pointing in an arbitrary
direction (the same as in Eq. (\ref{rhostate}));

\item each spin in the maximally mixed state.

\end{itemize}

The latter state is represented by the following operator:
\begin{eqnarray}
\tau_2 = \frac{1}{4}I^{\otimes2} =\frac{1}{4} \left(
|0\rangle\langle 0|+|1\rangle\langle 1| \right) \otimes \left(
|0\rangle\langle 0|+|1\rangle\langle 1| \right),
\end{eqnarray}
where $|0\rangle$ and $|1\rangle$ are any two orthogonal spin
states.

The strategy now is exactly the same as before. It relies on the
fact that if the state is $\rho_2$ the particles can, due to their
indistinguishability, give only one result (antibunching in the
case of fermions or bunching in the case of bosons), while if the
input state is $\tau_2$ both results are possible. This time, the
probability of success is:
\begin{equation}
P_{BS} (\rho_2,\tau_2)=\frac{5}{8},
\end{equation}
and this coincides with the Helstrom result
\begin{equation}
P_{H} (\rho_2,\tau_2) = \frac{5}{8},
\end{equation}
so our procedure is optimal in this case too. The calculations
leading to the above results are analogous to the ones in the
previous discrimination case.

\begin{figure}[ht]
\begin{center}
\epsfig{file=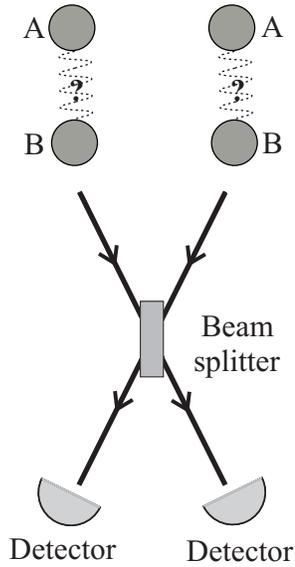, width=1.5in}
\end{center}
\caption{This figure represents the set-up for our entanglement
detection scheme. We have two equal pairs of identical particles
and consider their internal degrees of freedom. Both pairs are in
the same pure state: either separable or maximally entangled. We
take the same particle (for instance, $B$) from each pair and
interfere them at a $50/50$ beam splitter. If the states are
disentangled then the particles are indistinguishable and,
depending on the statistics, will either only bunch or only
antibunch. Otherwise, if the states are maximally entangled, the
particles are maximally mixed, meaning that they can be
(probabilistically) distinguished and hence the statistics does
not influence their behavior. This entanglement detection is a
particular instance of our state discrimination scheme discussed
in the paper.} \label{Fig. set-up}
\end{figure}

One of the most interesting applications of our approach can be
found in the detection of entanglement. We illustrate this in the
case of pure states of two particles. Suppose, for example, that
we are to discriminate any maximally entangled state of two qubits
from any disentangled (product) state. As before, the qubits are
supported by the internal degrees of freedom of two identical
particles (labelled, say, $A$ and $B$). Suppose in addition that
we are given two identical copies of the state. In order to detect
entanglement, we take the same particle (either $A$ or $B$) from
each pair and interfere them at a beam splitter, as shown in Fig.
\ref{Fig. set-up}. The crux of the argument is that if the state
is entangled, then the reduced states of these particles will be
maximally mixed as in the state $\tau_2$. On the other hand, if
the state is separable, then the two interfering particles are in
the state $\rho_2$. This is the same as in our discrimination
procedure above, and so there is a probability of $5/8$ to detect
entanglement. This example can be generalized to other entangled
pure states, and, more interestingly, to more particles. We note
that there is a close analogy between this method and our
entanglement concentration scheme in \cite{paunkovic}. In
particular, if the two states were less than maximally entangled,
then by detecting entanglement we would actually also amplify it
(see \cite{paunkovic} for more details).

Another interesting application is in mixed state purification, as
in \cite{artur}. Suppose that we start with two qubits, each in
some mixed state. We would like to make the state of these qubits
purer (in the sense of having lower linear entropy), but also to
preserve their original direction in the Bloch sphere. The optimal
way of doing so (as proven in \cite{artur}) is to project the
joint state onto the symmetric subspace, in which case the
resulting mixed state is purer and yet preserves the original
direction. If the projection is unsuccessful, the qubits are
thrown away. This is exactly the same as our probabilistic
discrimination with a beam splitter.

\begin{figure}[ht]
\begin{center}
\epsfig{file=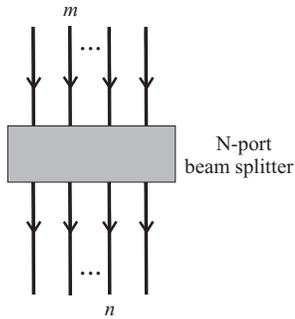, width=1.5in}
\end{center}
\caption{This diagram represents a multiport beam splitter with
$N$ inputs and $N$ outputs. The overall output state depends not
only on the input, but also on the statistics (either fermionic or
bosonic) of the identical particles involved. We have labelled two
arbitrary ports, the $m$-th input port and the $n$-th output
port.} \label{Fig. Multiport}
\end{figure}

We would now like to investigate the generalization of the above
results to $N$ particles. For this, we use a generalized (N-port)
balanced beam splitter, as shown in Fig. \ref{Fig. Multiport},
which acts only on the spatial degrees of freedom of the input
particles. This action is given by a unitary matrix $U_N$, with
elements:
\begin{equation}
u_{mn}=\frac{1}{\sqrt N}e^{i\frac{2\pi}{N}(m-1)(n-1)}.
\end{equation}
(Note that there exist alternative descriptions of balanced
multiport beam splitters \cite{zeilinger}.) The square of the norm
of each element in the matrix represents the probability that the
particle in the $m$-th input arm of the beam splitter ends up in
the $n$-th output arm. Since all these elements have norm
$\frac{1}{\sqrt N}$, we have a representation of a balanced N-port
beam splitter.

The aim is now to discriminate between the N-particle
generalizations of $\rho_2$ and $\tau_2$. Those states are given
by:
\begin{equation}
\rho_N=\frac{1}{4\pi}\int d\Omega
\left(|\Omega\rangle\langle\Omega|\right)^{\otimes N},
\end{equation}
and
\begin{equation}
\tau_N=\frac{1}{2^N}
\left(|0\rangle\langle0|+|1\rangle\langle1|\right)^{\otimes N}.
\end{equation}
To calculate the Helstrom probability, one has to diagonalize the
matrix $\rho_N-\tau_N$. This turns out to be straightforward once
we notice that $\rho_N$ can also be represented as an equal
mixture of all possible symmetric states within the basis $\{
|S_i\rangle : i=1,\ldots,N+1=|S|\}$ of $N$ qubits. Then, we have:
\begin{equation}
\rho_N=\frac{1}{|S|}\sum_{i=1}^{|S|} |S_i\rangle\langle S_i|.
\end{equation}
It is now easy to calculate the Helstrom formula by expanding
$\rho_N$ in a basis consisting of the union of a basis of the
symmetric sub-space and a basis of its orthogonal complement. The
result is:
\begin{equation}
\label{bf}
P_H (\rho_N, \tau_N) = 1-\frac{(N+1)}{2^{(N+1)}}.
\end{equation}
Alternatively, we can calculate the average probability of success
to distinguish states $\rho_N$ and $\tau_N$ using the following
expression:
\begin{equation}
\label{heuristic} P_H (\rho_N, \tau_N) = \frac{1}{2}1 +
\frac{1}{2}p.
\end{equation}
Here, the $1/2$ factors refer to the fact that the two states are
prepared with equal probability. The term $1$ comes from the fact
that the state $\rho_N$, supported on the symmetric subspace (of
dimension $d_S=N+1$), is always identified reliably as such, and
$p$ is the probability of identifying the other state, $\tau_N$.
Since $\tau_N$ is maximally mixed, it is uniformly distributed
over the whole space of $N$ qubits (of dimension $d=2^N$). In this
case $p=\frac{1-d_S}{d}$, which after substitution in Eq.
(\ref{heuristic}) gives Eq. (\ref{bf}) right away.

The problem to apply our discrimination scheme to $N$ particles is
that it becomes exponentially hard to calculate $P_{BS} (\rho_N,
\tau_N)$ as $N$ increases. Moreover, it is not clear which
inference strategy should be followed. For fermions, the natural
generalization seems to be to associate the antibunching results
with $\rho_N$ and the others with $\tau_N$. For bosons, on the
other hand, a more subtle strategy may be needed. This is because,
loosely speaking, for bosons there is no clear analogue of the
Pauli exclusion principle. Furthermore, even without having the
complete calculations for $N>2$, we would like to emphasize the
remarkable fact that the Helstrom probability $P_H (\rho_N,
\tau_N)$ is equal to the probability of success of a fermionic
beam splitter strategy described above, if calculated under the
assumption that the particles are classical (i.e., always
distinguishable by some arbitrary label), but obey a constraint
equivalent to the Pauli exclusion principle (not allowing more
than two particles in the same internal state to share the same
output arm of the beam splitter). The overall probability is then
calculated by summing up the probabilities of all the possible
outcomes rather than the amplitudes, as it would be done in the
quantum case. However, in the case of three fermions ($N=3$), we
have performed the full quantum calculations (i.e., taking
properly into account the effects of statistics) for a three-port
balanced beam splitter and obtained:
\begin{equation}
P_{BS}(\rho_3, \tau_3) =\frac{3}{4},
\end{equation}
which is equal to $P_H (\rho_3, \tau_3)$. We believe this result
of obtaining the optimal discrimination probability using the
effects of particle statistics (in multiports) can be generalized
to an arbitrary $N$, both for fermions and bosons, and we continue
research in this direction. For now this remains a conjecture. In
an optical lattice, with one particle in each lattice site, a
multiport beam splitter could probably be simulated by dissolving
N potential wells and then creating a new set of N wells
\cite{jaksch}. Of course, if the particles interact, then the
effective beam splitter will be modified, and here we only point
out the plausibility of creating multiport beamsplitters (or
multiparticle interference) in an optical lattice.

In this paper we have shown that it is possible to perform an
optimal quantum information processing task using only the effects
of particle statistics. In particular, we have presented a
strategy for discriminating between two non-orthogonal states of
two qubits (encoded in the internal degrees of freedom of
identical particles) using beam splitters. We have considered two
discrimination scenarios and in each of them our strategy differs
(symmetrically) between fermions and bosons, but offers the same
efficiency.  We also pointed out how our discrimination scheme can
be applied to detecting entanglement and purifying mixed states.
In addition, we have calculated the Helstrom probability for $N$
qubits in one of our discrimination scenarios. We have shown by
explicit calculation that this probability can be achieved in a
fermionic three-port beam splitter strategy and that it is the
same as the fermionic strategy for general $N$ if the fermions are
considered as classical particles that obey the Pauli exclusion
principle as the only additional constraint. An advantage of our
method is that it can also be easily implemented with the current
technology.

Our work suggests a number of interesting research directions. One
problem is to prove the optimality of the beam splitter strategy
in the case of $N$ qubits and its application to multiparty
entanglement detection. This, we hope, will answer the question of
weather the symmetry between fermions and bosons in our strategy
will be preserved for a generalized beam splitter. It may also
lead to a simple and physically intuitive selection principle
governing bosonic behavior. Another possible direction is to
classify all the pairs of states that can be optimally
discriminated with our scheme. Finally, our results suggest that
it would be worth to further explore the role of particle
statistics in quantum information tasks, in particular in
efficient quantum computation.

After the conclusion of this paper, an interesting related work
was pointed out to us \cite{barnett}. The authors would like to
thank G. Castagnoli on many useful discussions on the link between
computational complexity and particle statistics. His ideas have
stimulated much of the work in this paper. We also would like to
thank F. A. Bovino and P. Varisco for useful discussions on the
role of particle statistics in quantum information processing.
S.B. acknowledges support from the NSF under Grant Number
EIA-00860368. Y.O. acknowledges support from Funda\c{c}\~{a}o para
a Ci\^{e}ncia e a Tecnologia from Portugal and wishes to thank the
Institute for Quantum Information at Caltech for their
hospitality. N.P. and V.V. thank Elsag S.p.A. for financial
support and Centro de F\'isica de Plasmas at IST in Lisbon for
their hospitality. V.V. acknowledges support from Hewlett-Packard
company, EPSRC and the European Union projects EQUIP and TOPQIP.
This work has been partly supported by the QUIPROCONE
collaboration grant number 044.

%____________________________________________________________

\vspace{6cm}

%______________________________________________________________________ FIGURES

\end{document}